\documentclass[fleqn,usenatbib]{mnras}

\usepackage{newtxtext,newtxmath}
\usepackage[T1]{fontenc}

\DeclareRobustCommand{\VAN}[3]{#2}
\let\VANthebibliography\thebibliography
\def\thebibliography{\DeclareRobustCommand{\VAN}[3]{##3}\VANthebibliography}

\usepackage{graphicx}	% Including figure files
\usepackage{amsmath}	% Advanced maths commands
\title{A possible overall scenario for the outburst evolution of MAXI J1820+070 revealed by \textit{Insight}-HXMT}

\author[J. Q. Peng et al.]{
J. Q. Peng,$^{1,2}$\thanks{E-mail: pengjq@ihep.ac.cn}
 S. Zhang$^{1}$\thanks{E-mail: szhang@ihep.ac.cn},  Y. P. Chen$^{1}$\thanks{E-mail: chenyp@ihep.ac.cn}, L. D. Kong$^{1,2,3}$, P. J. Wang$^{1,2}$, S. N. Zhang$^{1,2}$,  L. Ji$^{4}$, L. Tao$^{1}$, 
\newauthor
J. L. Qu$^{1,2}$,   M. Y. Ge$^{1}$, Q. C. Shui$^{1,2}$, J. Li$^{5,6}$, Z. Chang$^{1}$, Z. S. Li$^{7}$, Y. X. Xiao$^{1,2}$
\\  
$^{1}$ Key Laboratory of Particle Astrophysics, Institute of High Energy Physics, Chinese Academy of Sciences, Beijing 100049, China\\
$^{2}$ University of Chinese Academy of Sciences, Chinese Academy of Sciences, Beijing 100049, China\\
$^{3}$ Institute f{\"u}r Astronomie und Astrophysik, Kepler Center for Astro and Particle Physics, Eberhard Karls, Universit{\"a}t, Sand 1, D-72076
T{\"u}bingen, Germany\\
$^{4}$ School of Physics and Astronomy, Sun Yat-Sen University, Zhuhai, 519082, China\\
$^{5}$ CAS Key Laboratory for Research in Galaxies and Cosmology, Department of Astronomy,
University of Science and Technology of China, Hefei 230026, China\\
$^{6}$ School of Astronomy and Space Science, University of Science and Technology of China,
Hefei 230026, China\\
$^{7}$ Key Laboratory of Stars and Interstellar Medium, Xiangtan University, Xiangtan 411105, Hunan, China\\
}

\date{Accepted XXX. Received YYY; in original form ZZZ}

\pubyear{2022}

\begin{document}
\label{firstpage}
\pagerange{\pageref{firstpage}--\pageref{lastpage}}
\maketitle

\begin{abstract}
We study the spectral and temporal properties of the black hole X-ray transient binary MAXI J1820+070 during the 2018 outburst  with  \textit{Insight}-HXMT observations. 
The outburst of MAXI J1820+070 can be divided into three intervals.
For the two intervals of the outburst, we find that low-energy (below 140 keV) photos  lag high-energy (140-170 keV) ones,
while in the decay of the outburst,   high-energy photons lag  low-energy photons, both with a time  scale  of the order of days. Based on these results, the canonical hysteresis effect of the  'q'  shape in the hardness-intensity diagram can be reformed into a roughly linear shape by taking into account the lag corrections between different energy bands.  Time  analysis shows that the high-frequency break of hard X-rays, derived from the power density spectrum of the first interval of the outburst is in general larger and more variable than that of  soft X-rays.  The spectral fitting shows that the coverage fraction of the hard X-rays drops sharply at the beginning of the outburst to around 0.5, then increases slightly. The coverage fraction drops to roughly zero once the source steps into  a soft state and  increases gradually  to unity when the source returns to a low hard state. We discuss the possible overall evolution scenario of corona  hinted from these discoveries.
\end{abstract}


%%%%%%%%%%%%%%%%%%%%%%%%%%%%%%%%%%%%%%%%%%%%%%%%%%

%%%%%%%%%%%%%%%%% BODY OF PAPER %%%%%%%%%%%%%%%%%%

\section{Introduction}

Black hole X-ray binaries (BHXRBs) are generally composed of a star and a central black hole. The central black hole accretes matter through the overflow of the Roche-lobe, and the accreted gas forms an accretion disk around the black hole \citep{1973Shakura}. Due to the viscous effect of the inner and outer layers of the gas, the angular momentum is transferred outward, and the material in the disk is accreted inward \citep{1974Lynden-Bell, 2002Frank}. 
The gravitational energy of the accreted material is converted into the thermal emission of the disk, non-thermal emission from corona and jet, caused by the Compton scattering of the hot electron cloud of the corona/jet off the hot photons of the disk \citep{2007Done}.

BHXRBs are classified into persistent or transient sources \citep{2016Tetarenko, 2018Sreehari}.
The persistent sources are always active and accrete the companion star matter with a high accretion rate, and hence remain bright for a long time \citep{2009Deegan}. 
The transient sources are in quiescent state  most of the time, and rarely accrete material from their companion stars. When they  enter the outburst state from the quiescent state, thermal instability and viscous instability will increase the accretion rate and lead to bright X-ray emission \citep{1995Cannizzo, 2001Lasota, 2016Corral-Santana}. The spectral states of the black hole transient sources can be divided into Low/Hard States (LHS), High/Soft States (HSS), and Intermediate States (IMS) during the outburst \citep{2005Belloni}.
In the hard state, the non-thermal emission is mainly the high-energy emission caused by the Comptonization of hot electrons \citep{1975Thorne}.
In the soft state, the  spectrum is dominated by the thermal emission of the accretion disk, and  the spectrum has a typical photon index $\Gamma$ $\sim$ 3. In addition to thermal emission,  a power-law tail can be seen occasionally \citep{1996Tanaka}. The intermediate state  is seen during transitions between the low and high states \citep{2005Belloni, 2005Homan}.

The trajectory of a typical outburst of the black hole X-ray transient sources on the hardness-intensity diagram (HID) is 'q' \citep{2001Homan, 2004Fender, 2012Motta}, and different spectral states have different positions on the HID during an outburst, showing the hysteresis between rising and decay, e.g. similar spectral state transitions occur at different flux \citep{2005Meyer-Hofmeister}. This hysteresis exists in most black hole binary systems \citep{2003Maccarone}.
For the explanation of 'q' diagram and hysteresis effect,
\cite{weng2021} presented a detailed time-lag analysis of a  BHXRB, MAXI J1348--630, which was  intensively monitored by \textit{Insight}-HXMT over a broad energy band (1--150 keV). They concluded that the observed time-lag between radiations of the accretion disk and the corona leads naturally to the hysteresis effect in the 'q'-diagram. However, it is unclear whether this mechanism universally works for other BHXRBs and an outburst in the decay phase.

 MAXI J1820+070 (ASASSN--18ey) is a  black hole low-mass X-ray binary (LMXRB) discovered by the All-Sky Automated Survey for SuperNovae (ASAS-SN) \citep{2018Denisenko}. A week after ASAS-SN discovered ASASSN--18ey as an optical transient \citep{2018Tucker}, it was detected as an X-ray transient by MAXI/GCS  \citep{2018Kawamuro}.
 \cite{2020Atri} used VLBA and VLBI to measure the parallax of radio of MAXI J1820+070 accurately, and measured a distance to the source of 2.96 $\pm$ 0.33 kpc and a jet interpolation angle of 63 $\pm$ 3\textdegree.  The mass estimate of the central black hole is $8.48^{+0.79}_{-0.72}$ $M_{\odot}$ \citep{2020Torres}.
The spin of  MAXI J1820+070 is inferred as $0.2^{+0.2}_{-0.3}$ by adopting a continuum fitting method with \textit{Insight}-HXMT observations \citep{2021Guan}, 
and  $0.799^{+0.016}_{-0.015}$ by fitting the Relativistic Precession Model (RPM) with the Neutron Star Interior Composition Interior Explorer (\textit{NICER})  observations \citep{2021Bhargava}.
In the LHS, \cite{Kara2019} applied a reverberation mapping upon \textit{NICER} data and found that the timescale of the reverberation lags shortens by an order of magnitude over a period of weeks, while the broad iron line profile remains constant. They explained that  the corona  vertically contracted and the disk was not truncated. These two hypothesizes are supported in \cite{2019Buisson}. However, other studies indicated  that the disk was truncated in the hard state \citep{2021Axelsson, 2021Zdziarski, 2021DeMarco, 2022Zdziarski}.
\cite{2021Ma} discovered low-frequency quasi-periodic oscillations (LFQPOs) above 200 keV with  \textit{Insight}-HXMT data, and the phase lag of LFQPO was soft lag above 30 keV, which gradually increased with energy and reached $\sim$ 0.9 s in the 150--200 keV band. They concluded that the LFQPO probably originated from the precession of a small-scale jet.
Observations from \textit{Insight}-HXMT revealed that the gradually decreasing reflection component in the energy spectrum can be understood in a jet-outflowing scenario. When the corona was closer to the black hole, the outflow velocity was greater \citep{You2021}.
During the hard-to-soft transition, the spectral-timing analysis with \textit{NICER} data showed that the corona may have expanded \citep{wang2021}.

In this paper, we analyze  the entire outburst data of MAXI J1820+070 collected by  \textit{Insight}-HXMT and study the hysteresis effect and the evolution of the corona during the  outburst. In Section \ref{obser}, we describe the observations and data reduction. The detailed results are presented in Section \ref{result}. The results are discussed  and the  conclusions are presented in Section \ref{dis}.

\section{Observations and Data reduction}
\label{obser}

\textit{Insight}-HXMT  is the first Chinese X-ray astronomy satellite, which was successfully launched on 2017 June 15 \citep{2014Zhang, 2018Zhang, 2020Zhang}.  It carries  three scientific payloads: the low energy X-ray telescope (LE, SCD detector, 1–-15 keV, 384 $\rm cm^{2}$,
\citealt{2020Chen}), the medium energy X-ray telescope (ME, Si-PIN detector, 5--35 keV, 952 $\rm cm^{2}$, \citealt{2020Cao} ), and the high energy X-ray telescope (HE, phoswich NaI(CsI), 20--250 keV, 5100 $\rm cm^{2}$, \citealt{2020Liu}).

\textit{Insight}-HXMT carried out a Target of Opportunity (ToO) observation on 2018 March 14, and monitored the whole outburst  more than 140 times with a total exposure of 2560 ks accumulated from 2018 March to October. We extract the data from all three payloads using the \textit{Insight}-HXMT Data Analysis software {\tt{HXMTDAS v2.04}}. 
The data are filtered with the  criteria recommended of the \textit{Insight}-HXMT Data Reduction Guide {\tt v2.05}\footnote[2]{{http://hxmtweb.ihep.ac.cn/SoftDoc/648.jhtml}}

We subtract off the background in order  to obtain the net count rate for plotting the light curves: 1--5 keV and 5--10 keV for LE,  10--20 keV and 20--30 keV  for ME, and  20 to 170 keV with  intervals of 30 keV for HE (Figure \ref{lcurve}).
{\tt Xspec v12.12.1}\footnote[3]{{https://heasarc.gsfc.nasa.gov/docs/xanadu/xspec/index.html}} is used to perform analysis of spectrum, where  LE data above 2 keV are chosen to  suppress possible source pollution at lower energies, and power spectra for LE in 1--10 keV and HE in 20--50 keV are also produced.  One percent  systematic error is added to data, and errors are estimated via  Markov Chain Monte-Carlo (MCMC) chains with a length of 20000.

\section{Results}
\label{result}

\subsection{\textit{Insight}-HXMT light curve and time-lag}
\label{light curve}

\begin{figure*}
	\centering
	\includegraphics[angle=0,scale=0.35]{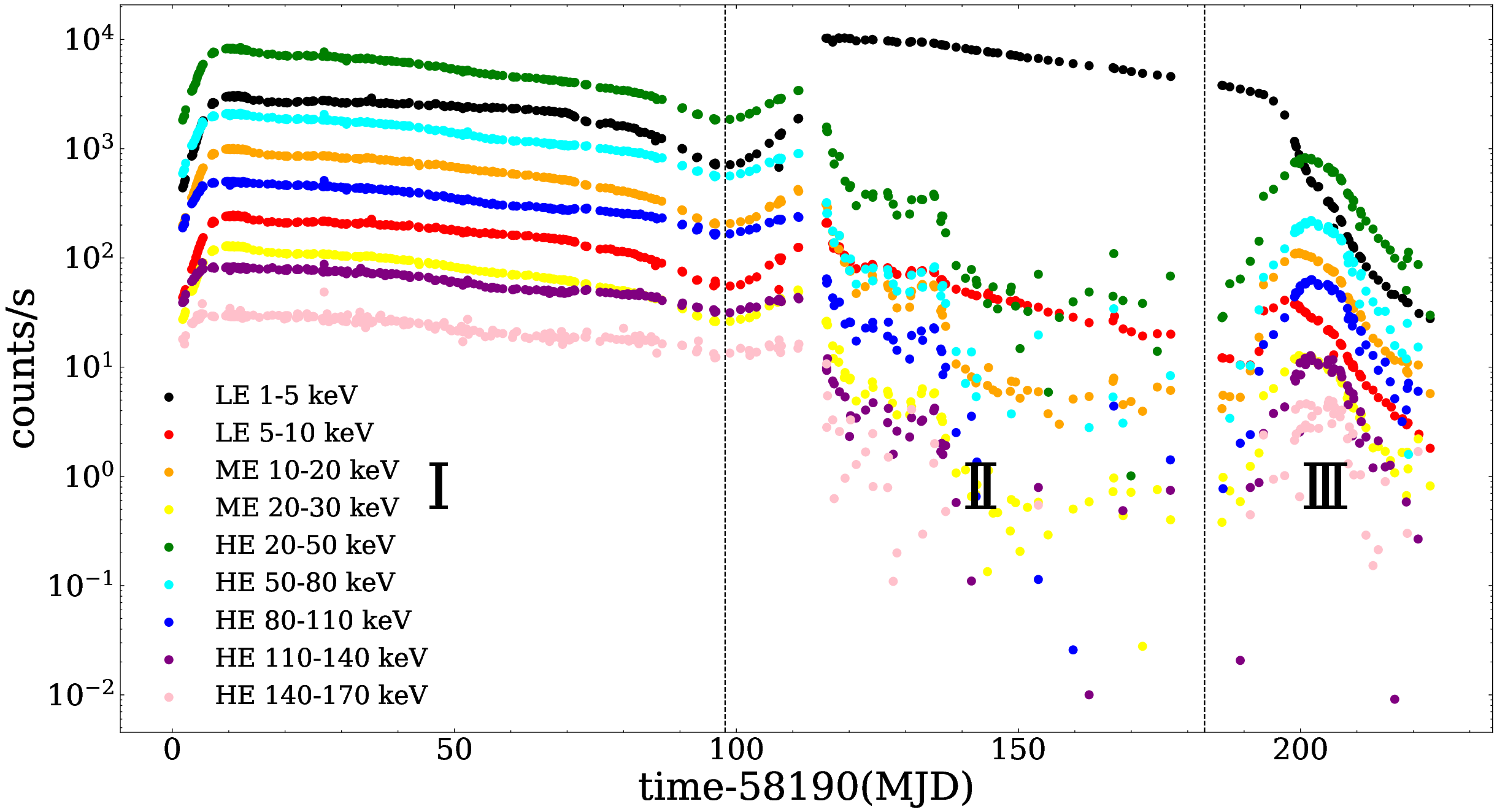}
	\caption{Light curves during the  outburst  of MAXI J1820+070 recorded by \textit{Insight}-HXMT in 2018.  \textit{Insight}-HXMT data are divided into nine different energy segments covering an energy band of 1--170 keV.
	For a better visual effect of  these light curves, we multiply these light curves by different constants.  
	The whole light curve of the   outburst  is divided into three zones: \uppercase\expandafter{\romannumeral1}, \uppercase\expandafter{\romannumeral2} and \uppercase\expandafter{\romannumeral3}.}
	\label{lcurve}
\end{figure*}

\begin{figure}
	\centering
%	\hspace{-1.1cm}
%	\flushleft
	\includegraphics[angle=0,scale=0.19] {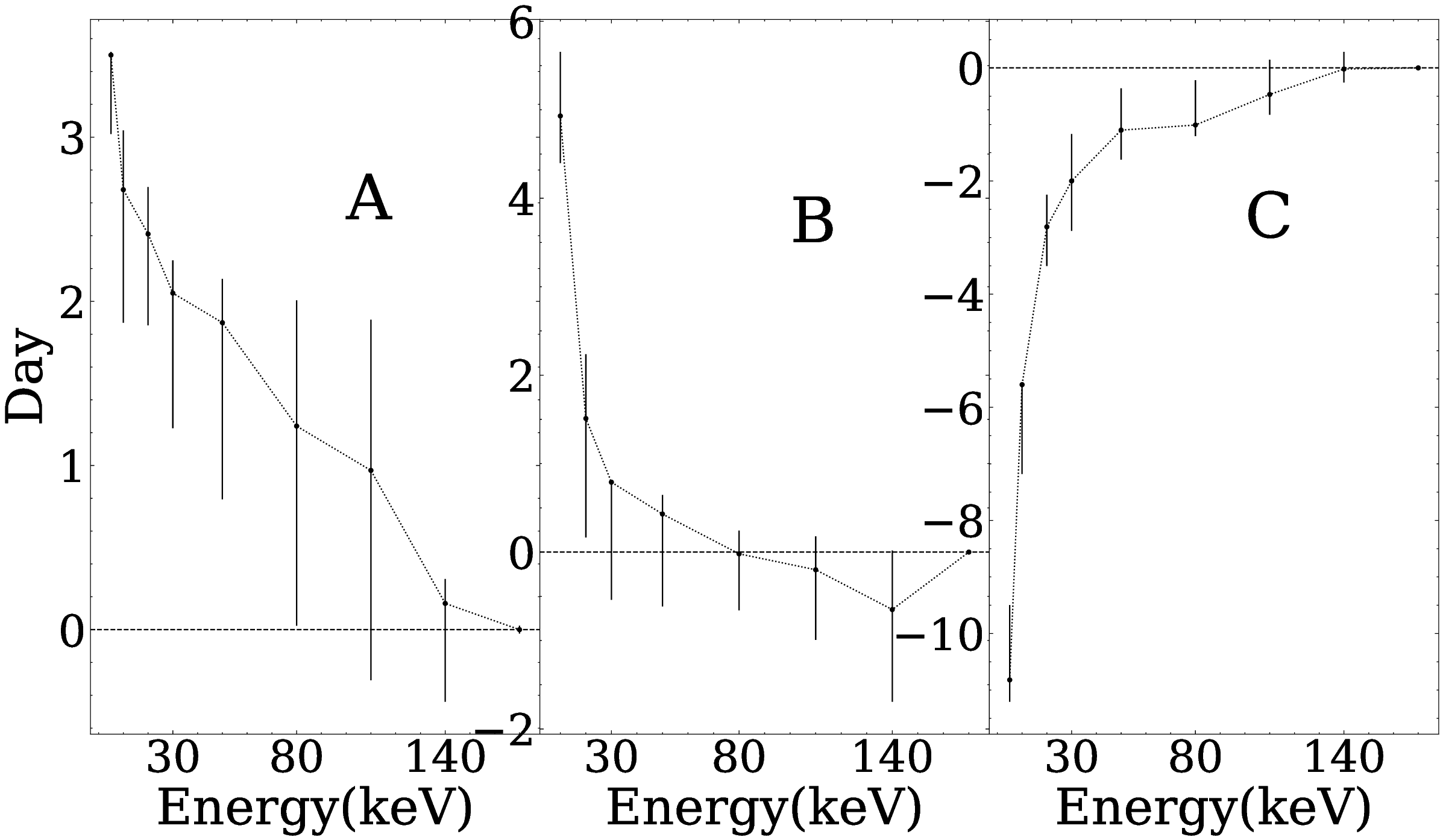}
    %\vspace{-0.3cm}
	\caption{The lag-energy relationship estimated  by the light curve of each energy segment with respect to HE 140--170 keV. A, B and C correspond to the three zones \uppercase\expandafter{\romannumeral1} , \uppercase\expandafter{\romannumeral2} and \uppercase\expandafter{\romannumeral3} in Figure \ref{lcurve}, respectively. The positive/negative lags denote the delay/proceed of the soft X-rays with respect to the hard X-rays.}
	
	\label{lag}
\end{figure}

 \textit{Insight}-HXMT began to observe MAXI J1820+070 on 2018 March 14  and  ended on  2018  October  21 \citep{2020Wang}.
 The whole outburst of MAXI J1820+070 was recorded
 The outburst of MAXI J1820+070 can be divided into three intervals.
 During the first interval of the outburst observed
 by \textit{Insight}-HXMT from  2018 March 14 (MJD 58191), the flux reached a peak  quickly and then decayed slowly. The end of the first interval of the outburst  on roughly 2018 June 17 (MJD 58286) was succeeded by the onset of the  second interval of the outburst  on 2018 June 19 (MJD 58288), shaping itself again in the form of 'rapid rise and slow decay'. In the second interval of the outburst,  MAXI J1820+070 entered the soft state, dominated by the soft component, where the count rate at hard X-rays is rather low.
In the decay, the hard X-ray flux increased and then  decayed after reaching the peak, along with the continuous decrease of the soft X-ray flux. Accordingly,  the whole light curves are divided into three zones: I, II and III, as shown in Figure \ref{lcurve}.

For each zone, we estimate the time lag of the light curves relative to the highest energy band (HE 140--170 keV)  using the interpolated cross-correlation method \citep{1998Peterson, 2018Sun, 2020Cai}. 
In the first interval of the outburst, the soft X-ray emission reached to peak  later than the hard X-rays, thus forming the so-called  soft lag.
As shown in Figure \ref{lag}, in zone I, time lag evolves almost linearly with decreasing energies, and has a maximum of about 3.5 days at the lowest energy band 1--5 keV. 
There exist as well soft lags in zone II, which manifest in a different manner with energies: a lag of about 5 days is only significantly present in the lowest energy band (1--5 keV). In zone III, the outburst evolves completely differently from zones I and II: instead of soft lag, one sees obvious hard lags, which increase with energies and reach a maximum of roughly 10 days with respect to the light curve of 1--5 keV (Figure \ref{lag}).

\subsection{Hardness-intensity diagram}
\label{HID}

\begin{figure*}
	\centering
%	\hspace{-1.1cm}
%	\flushleft
	\includegraphics[angle=0,scale=0.35] {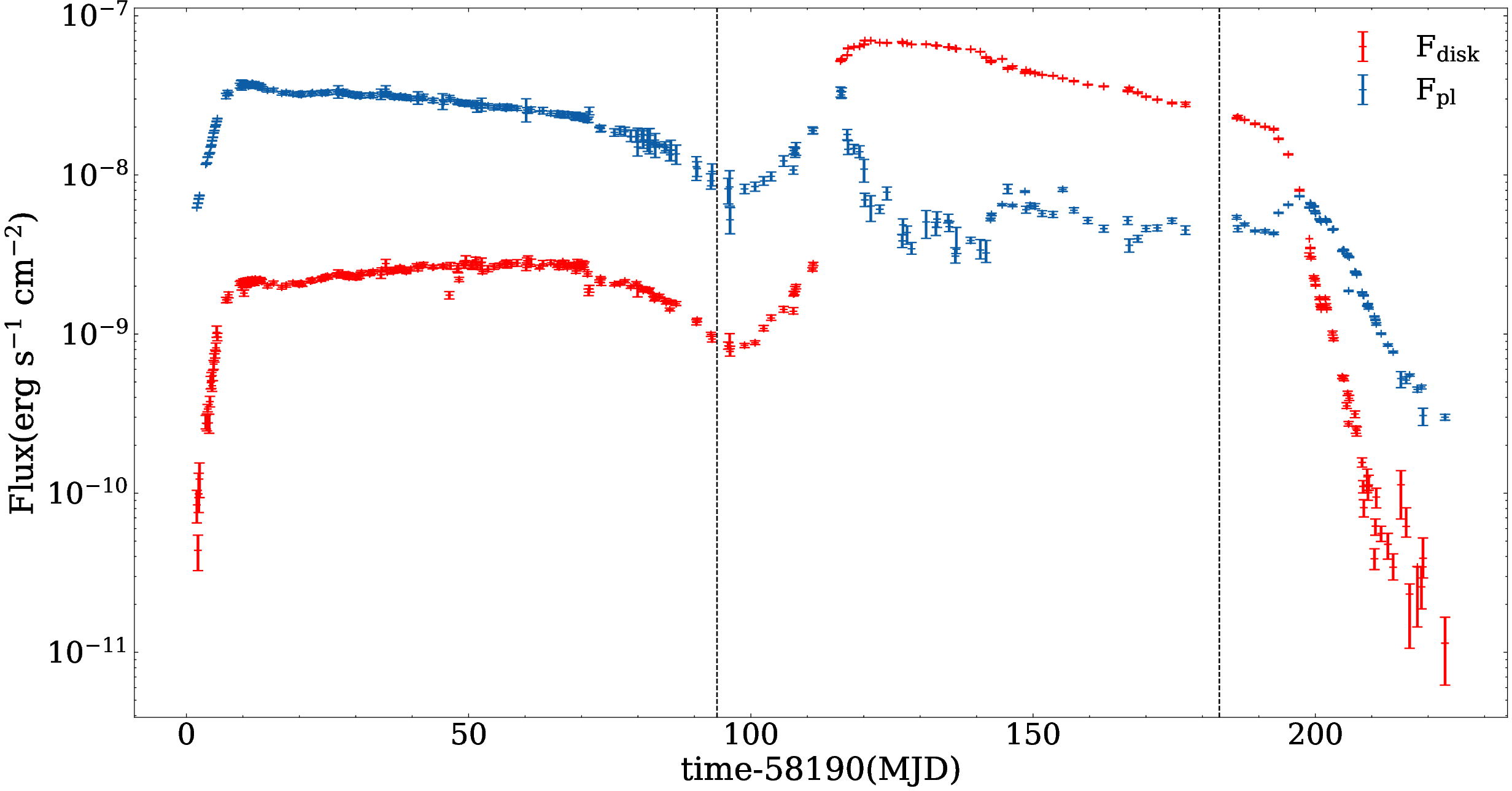}
    %\vspace{-0.3cm}
	\caption{ Evolutions of disk thermal component (Blue Cross) and the non-thermal component (Red Cross) of MAXI J1820+070, in 1--10 keV.}
	\label{flux}
\end{figure*}

\begin{figure}
	\centering
\begin{minipage}{0.5\textwidth}
  \includegraphics[angle=0,scale=0.2]{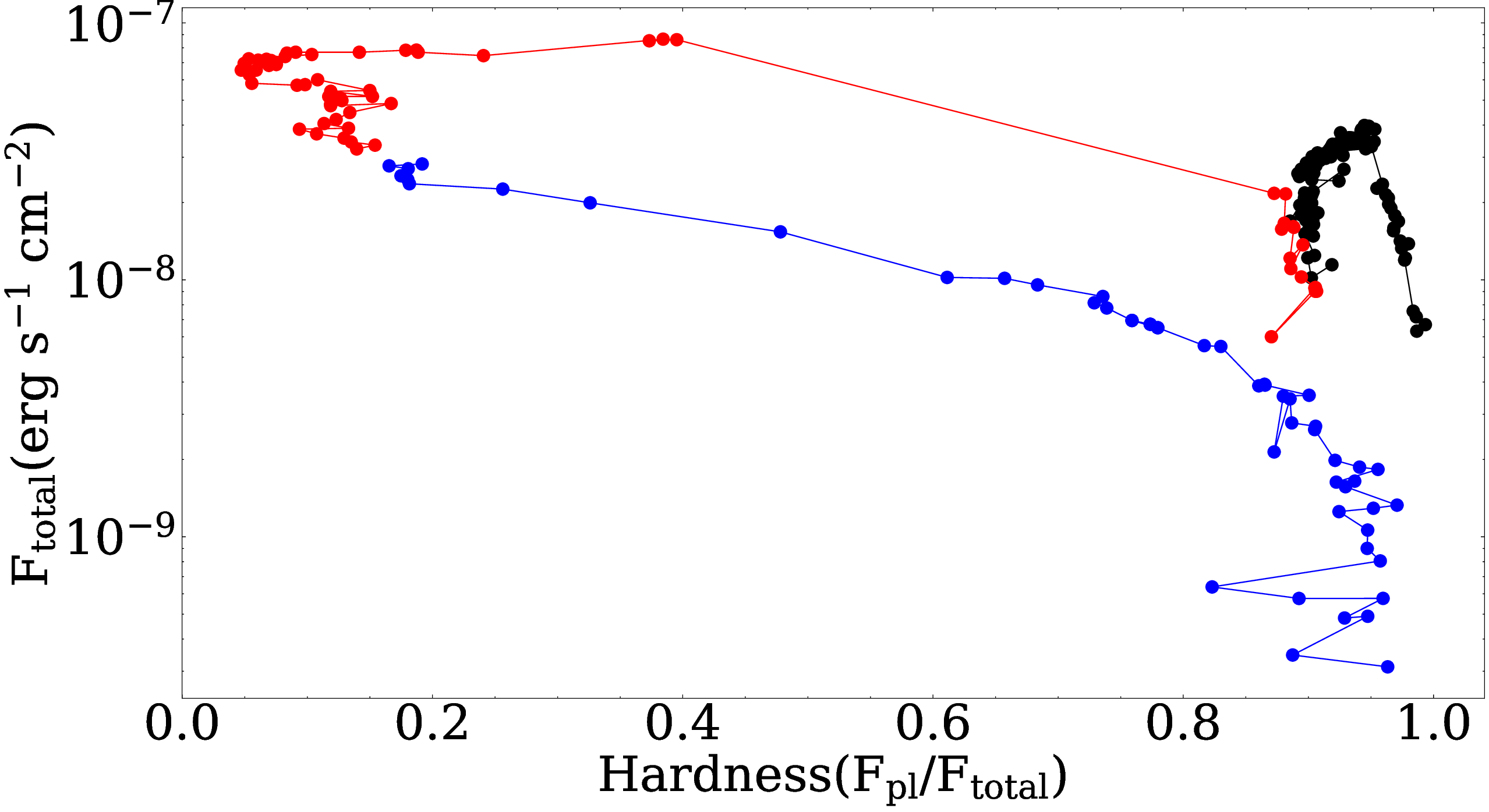} \\
  \includegraphics[angle=0,scale=0.2]{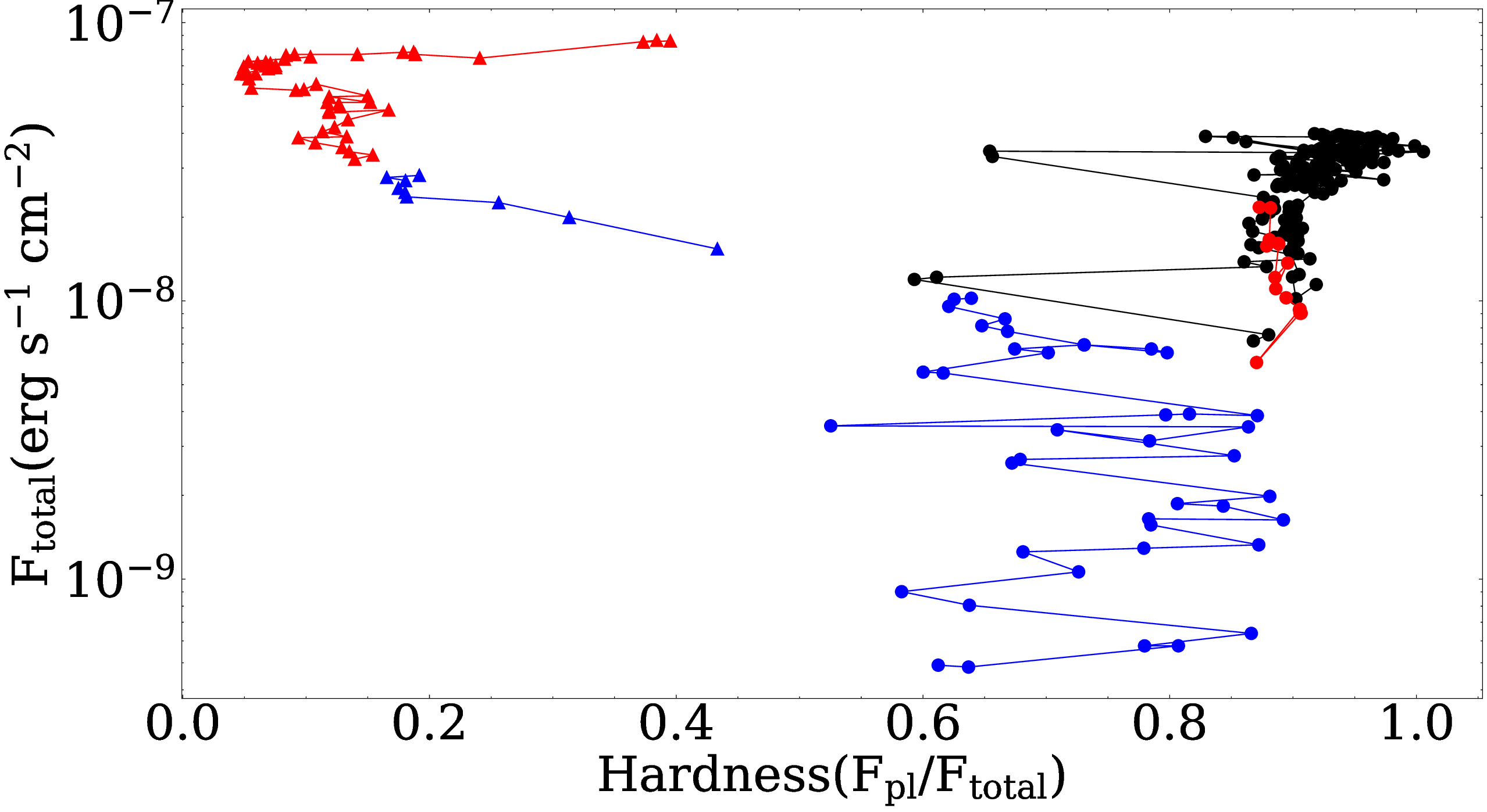}
  \caption{Top panel: the \textit{Insight}-HXMT hardness-intensity diagram of MAXI J1820+070, where the hardness is defined as the ratio of the power-law flux, $ F_{\rm pl}$, to the disk flux, $ F_{\rm disk}$, in 1--10 keV. The black, red and blue dots correspond to the first interval, the second interval and the decay of the outburst,  i.e. zones I, II and III in Figure \ref{lcurve}. Bottom panel: by taking into account the time lags, the tracks can be reformed mostly in a linear shape, except for the data in HSS.}
  \label{uncorrected HID}
  
\end{minipage}
\end{figure}

 We fit the  spectrum to obtain the unabsorbed flux of 1--10 keV (Figure \ref{flux}), and then construct the Hardness-Intensity Diagram (HID) of MAXI J1820+070 (See Section \ref{parameters} for details of the  spectrum fitting). Figure \ref{uncorrected HID} (top panel) shows the \textit{Insight}-HXMT HID of MAXI J1820 + 070, with the hardness ratio  defined as the ratio of the power-law flux to the total flux, in 1--10 keV. The HID  has  a 'q' shape typical to the  black hole transients. 
 The black, red and blue dots stand for the three zones as clarified in Figure \ref{lcurve}. 
 This HID shows that, zone I covers the first interval of the outburst when the source stayed  in a low-hard state (black dots in Figure \ref{uncorrected HID}), the transition from the end of the first interval of the outburst to the soft state of the second interval of the outburst is denoted by the red dots covered in zone II, and  the recovered  low-hard state as represented by the blue dots in zone III.

In order to investigate the effect of time lag on HID,  we estimate  the time lag between  the non-thermal component, $ F_{\rm pl}$, and the thermal component, $F_{\rm disk}$,  through cross-correlation \citep{1998Peterson, 2018Sun, 2020Cai}. It turns out  that $ F_{\rm disk}$ lags $F_{\rm pl}$  $\sim$ 1.52 days in zone I  and $\sim$ 6.8 days in zone II, and    $F_{\rm pl}$ lags $F_{\rm disk}$ $\sim$ 9.1 days in zone III. These time lags are in general consistent with those estimated with count rates in different energy bands  in Figure \ref{lag}. 

We correct the time lag between $ F_{\rm pl}$ and $F_{\rm disk}$ and align the peaks of $ F_{\rm pl}$ and $F_{\rm disk}$ in the three zones respectively. Accordingly, the  HID is modified from the original 'q' shape in the top panel of Figure \ref{uncorrected HID} to  a linear relationship, in the bottom panel of Figure \ref{uncorrected HID} for most spectral states except HSS.

\subsection{Break frequency}
\label{break frequency}

\begin{figure}
	\centering
    \begin{minipage}{0.5\textwidth}
        \centering
        \includegraphics[angle=0,scale=0.3]{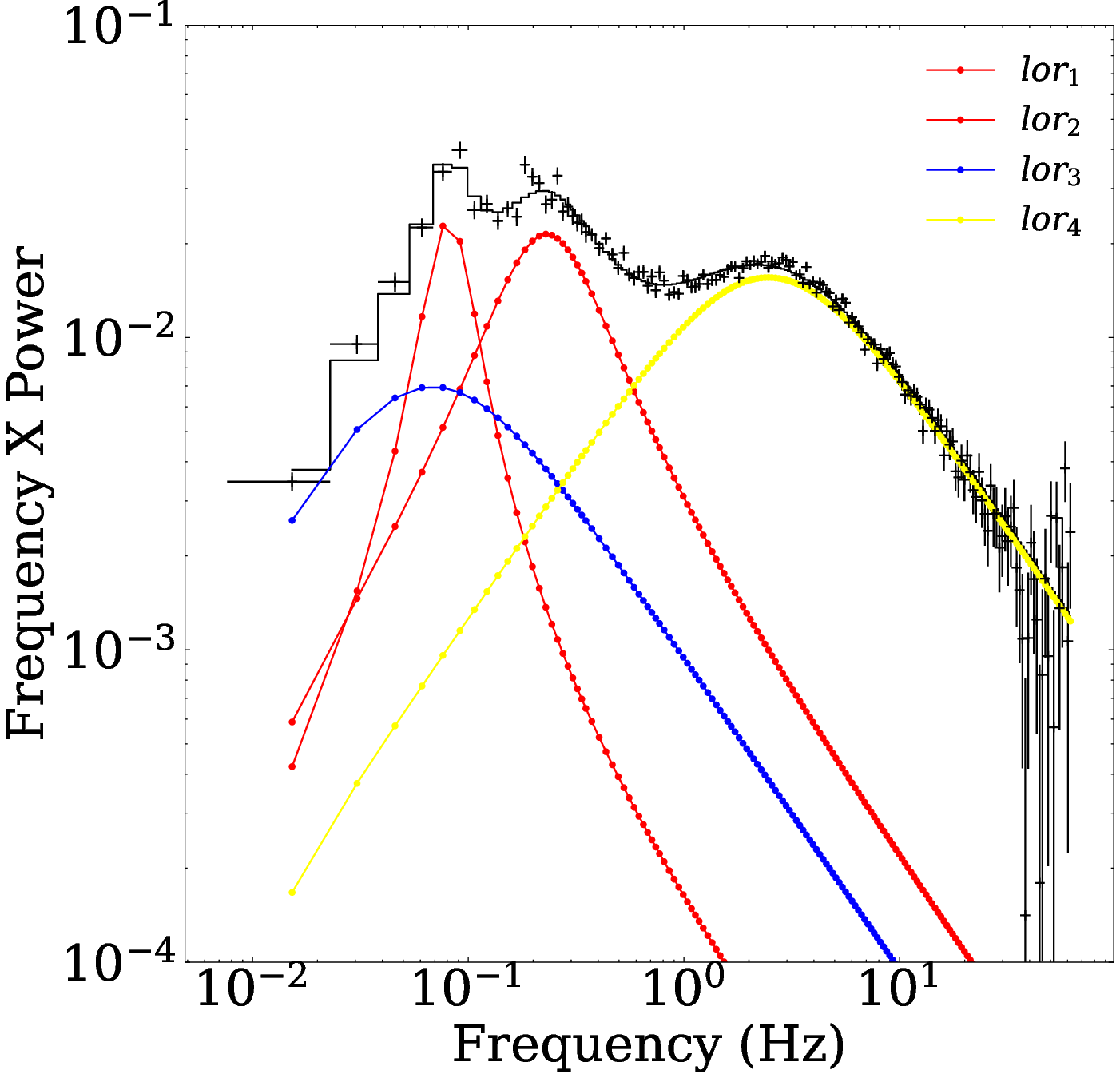} \\
        \includegraphics[angle=0,scale=0.3]{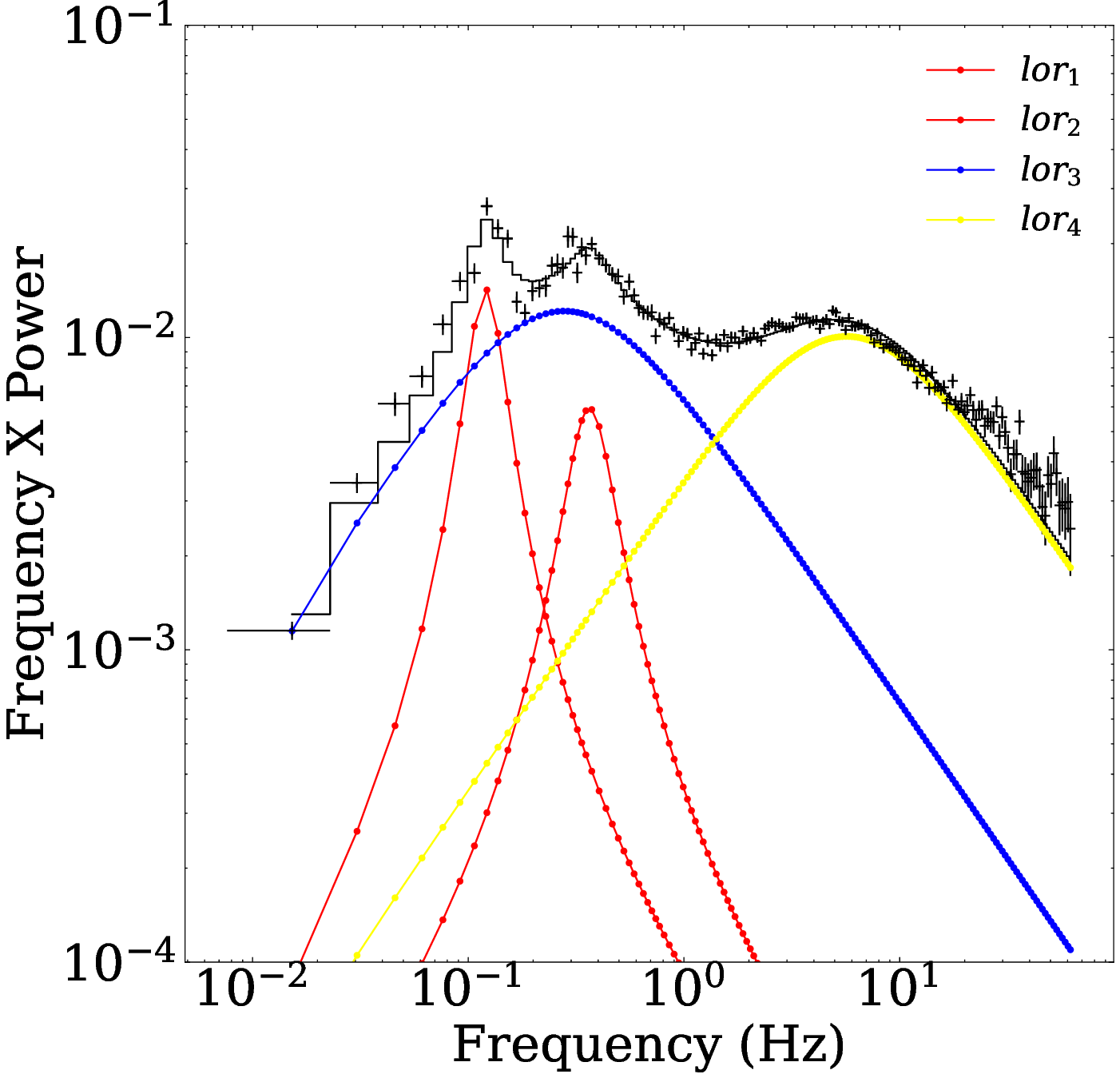}
        \caption{A representative PDS  of LE (1--10 keV) and HE (20--50 keV) fitted with multiple-Lorentzian. }
        \label{POWER}
    \end{minipage}
\end{figure}

\begin{figure}
	\centering
	\includegraphics[angle=0,scale=0.2] {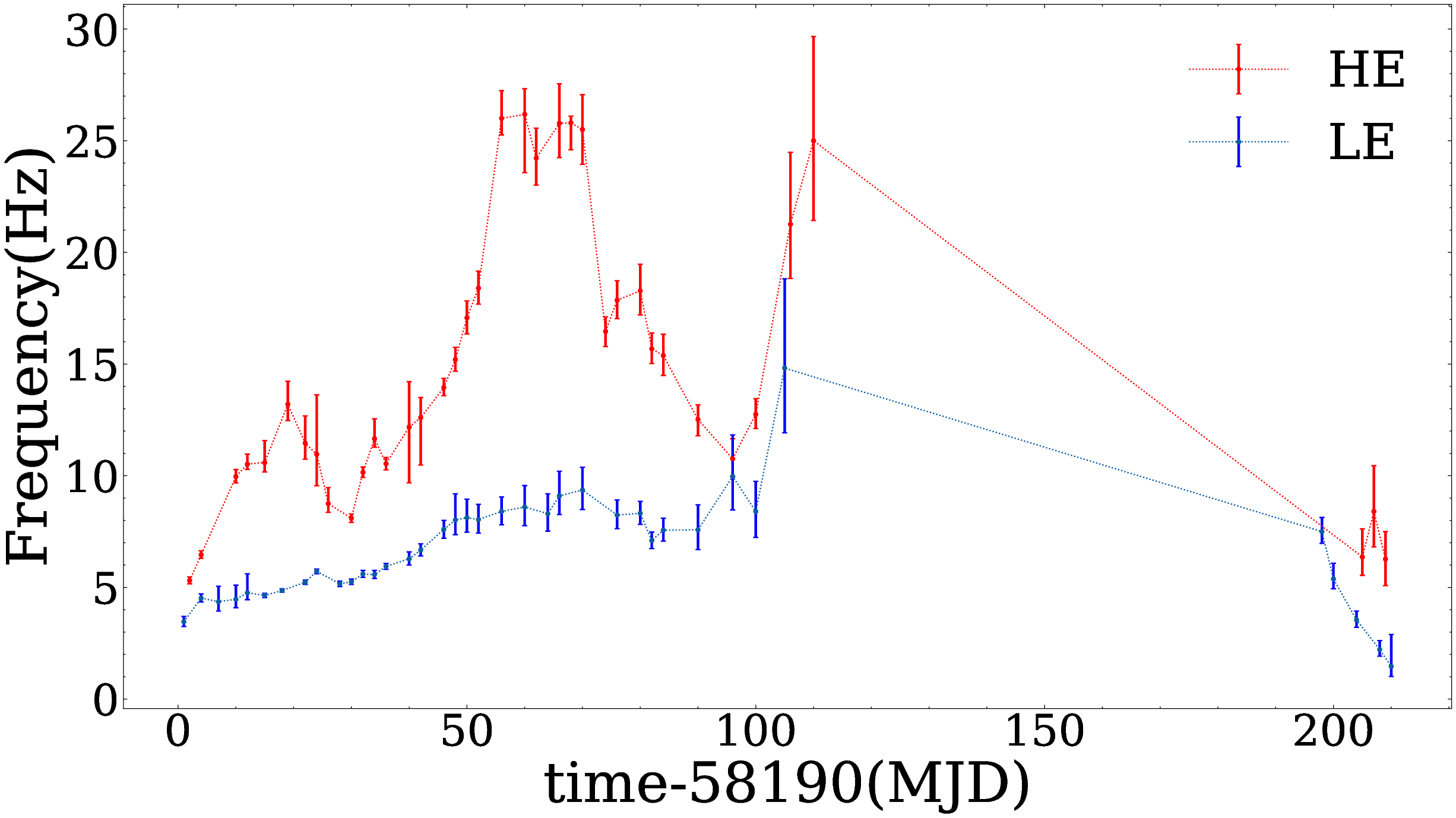}
	\caption{break frequency of PDS of HE in 20-50 keV and LE in 1-10 keV. Red and blue data correspond to break frequencies of $\rm Lor_{4}$  of HE and LE, respectively.}
	\label{frequency}
\end{figure}

To improve the statistics of the power density spectrum (PDS), we merge the light curves of LE (1--10 keV) and HE (20--50 keV)  by day, respectively. We then use the tool {\tt{powspec}}\footnote[1]{{https://heasarc.gsfc.nasa.gov/docs/xanadu/xronos/examples/powspec.html}} of {\tt{HEASOFT}}  to generate power spectra with  a time resolution of 0.008 s, the
interval of 65.5 s and Nyquist frequency of 62.5 Hz, under the Miyamoto normalization \citep{1990Belloni, 1992Miyamoto}.
The power spectra of black hole sources can be described by a sum of Lorentzian components \citep{2002Belloni}, as shown in Figure \ref{POWER}.

The characteristic frequency can be calculated as,  $\nu_{\rm max}$=$\sqrt{\nu^{2}_{0}+({\Delta/2})^{2}}$ \citep{1997Belloni}, where $\nu_{0}$ is the centroid frequency and $\Delta$ is the full width at half maximum (FWHM) of the Lorentzian function, \textit{Q} $\equiv$ $\nu_{0}$/$\Delta$.
Generally, the narrow component (\textit{Q}>2) is called quasi-periodic oscillations (QPO), and the wide component with low coherence is called noise.  We adopt Lorentzian functions to fit QPO with central frequency free and the noise with center frequency set to zero.

  As shown in Figure \ref{POWER}, the PDS has two broad components, $\rm Lor_{3}$ and $\rm Lor_{4}$, that could be fitted with zero-centered Lorentzians. Their characteristic frequency of $\rm Lor_{4}$ is the high  break frequency.
As shown in Figure \ref{frequency}, 
the break frequency of $\rm Lor_{4}$  in hard X-rays derived from the power spectra of the first interval of the outburst is in general higher and more variable than that in the soft X-rays: while the latter increases more or less monotonously, the break frequency of $\rm Lor_{4}$ in hard X-rays increases to form a sharp peak at around MJD 58260  and then  decreases toward the start of the second interval of the  outburst.  The PDS is rather weak when the source stays in the soft state, and thus the break frequency   can not be estimated in most parts of  Zone II and Zone III.  When the outburst returns to the hard state in the decay phase of the  outburst (zone III), the break frequency of $\rm Lor_{4}$  appears to  decrease for both energy bands. 

\subsection{Evolution of the Spectral parameters}
\label{parameters}

\begin{figure}
	\centering
	\begin{minipage}{\linewidth}
	    \centering
    	\includegraphics[angle=0,scale=0.28] {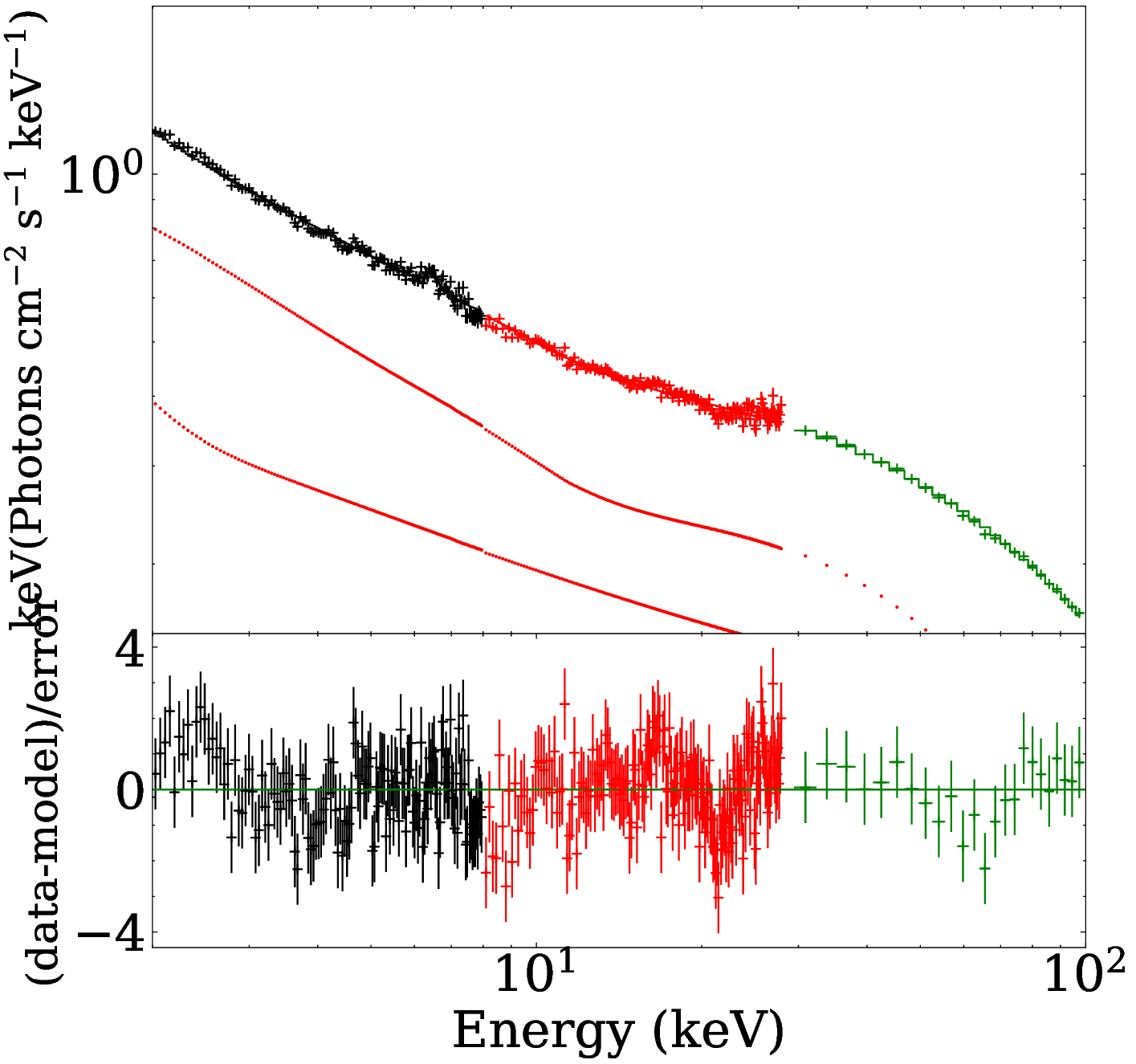}\\
    	\includegraphics[angle=0,scale=0.28] {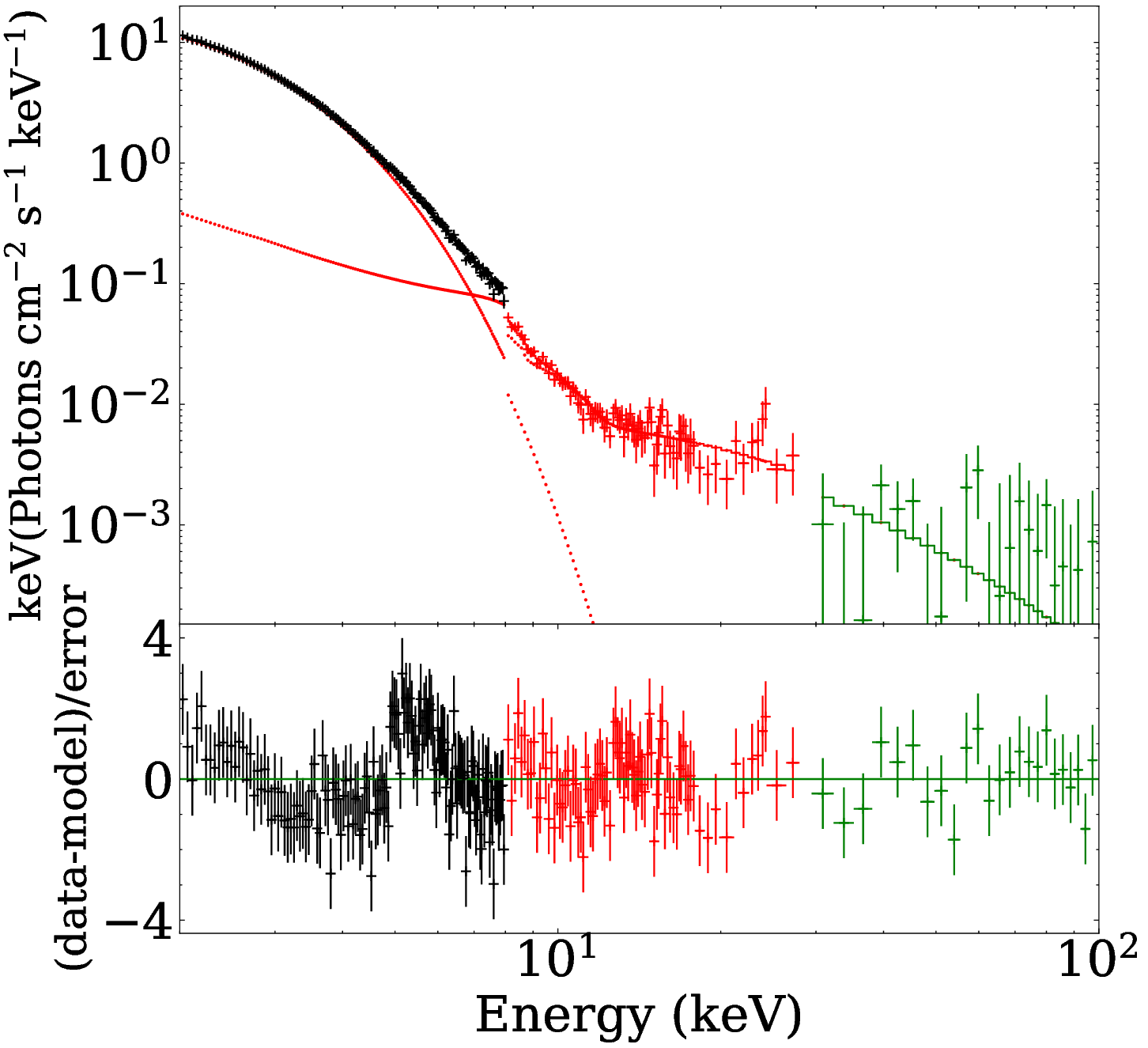} \\
    	\includegraphics[angle=0,scale=0.28] {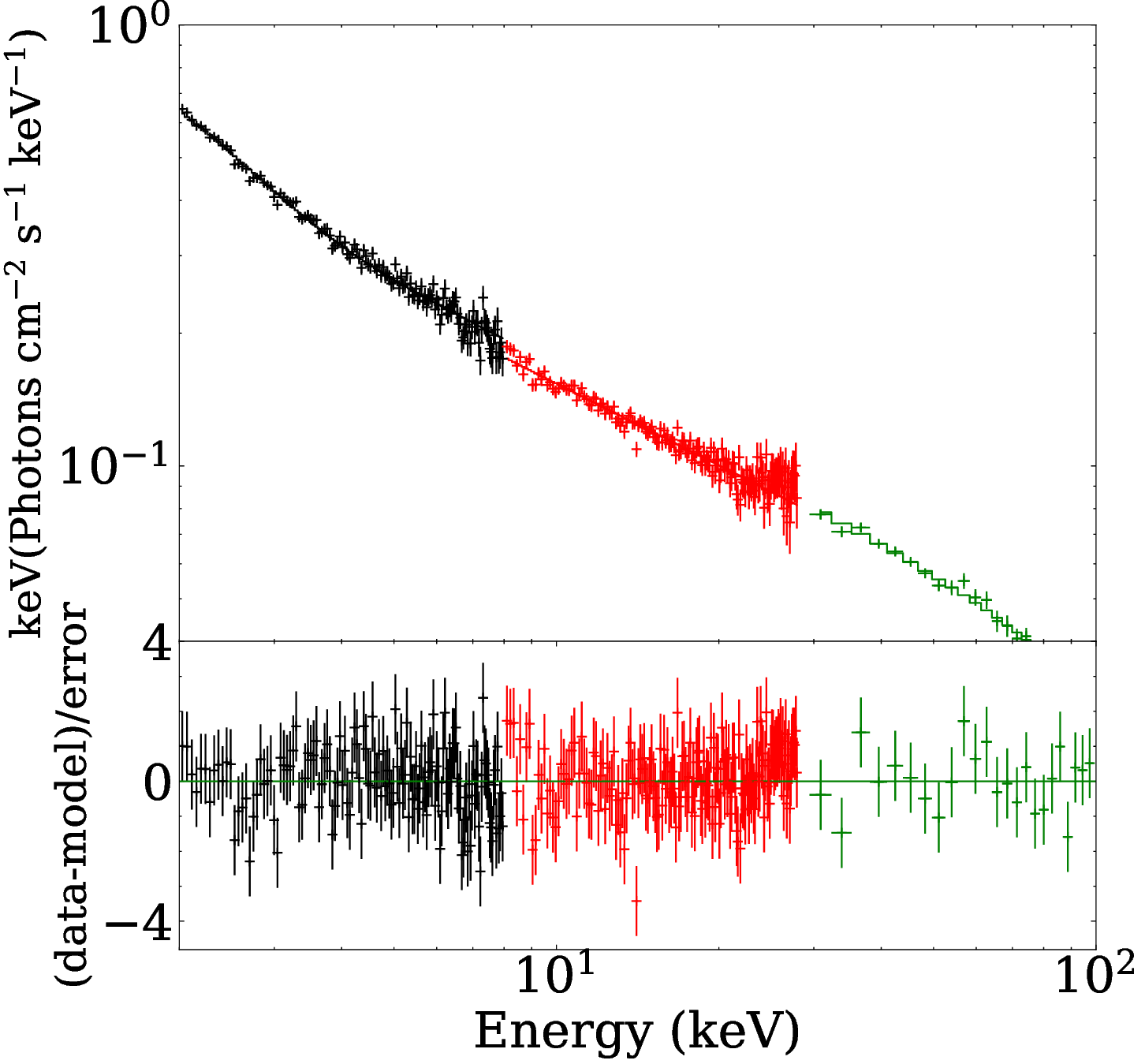}
    	\caption{ Spectral fittings for the data born out of Insight-HXMT observations of MAXI J1820+070 carried out in  ObsID P0114661100201, ObsID P0114661111101, ObsID P011466113501. Black, red and green symbols correspond to the LE, ME and HE data of \textit{Insight}-HXMT, respectively. The panel from top to bottom corresponds to the energy spectra taken in zones I, II, and III, respectively.  }
    	\label{spectrum1}
	\end{minipage}
\end{figure}

\begin{figure*}
	\centering
	\includegraphics[angle=0,scale=0.25] {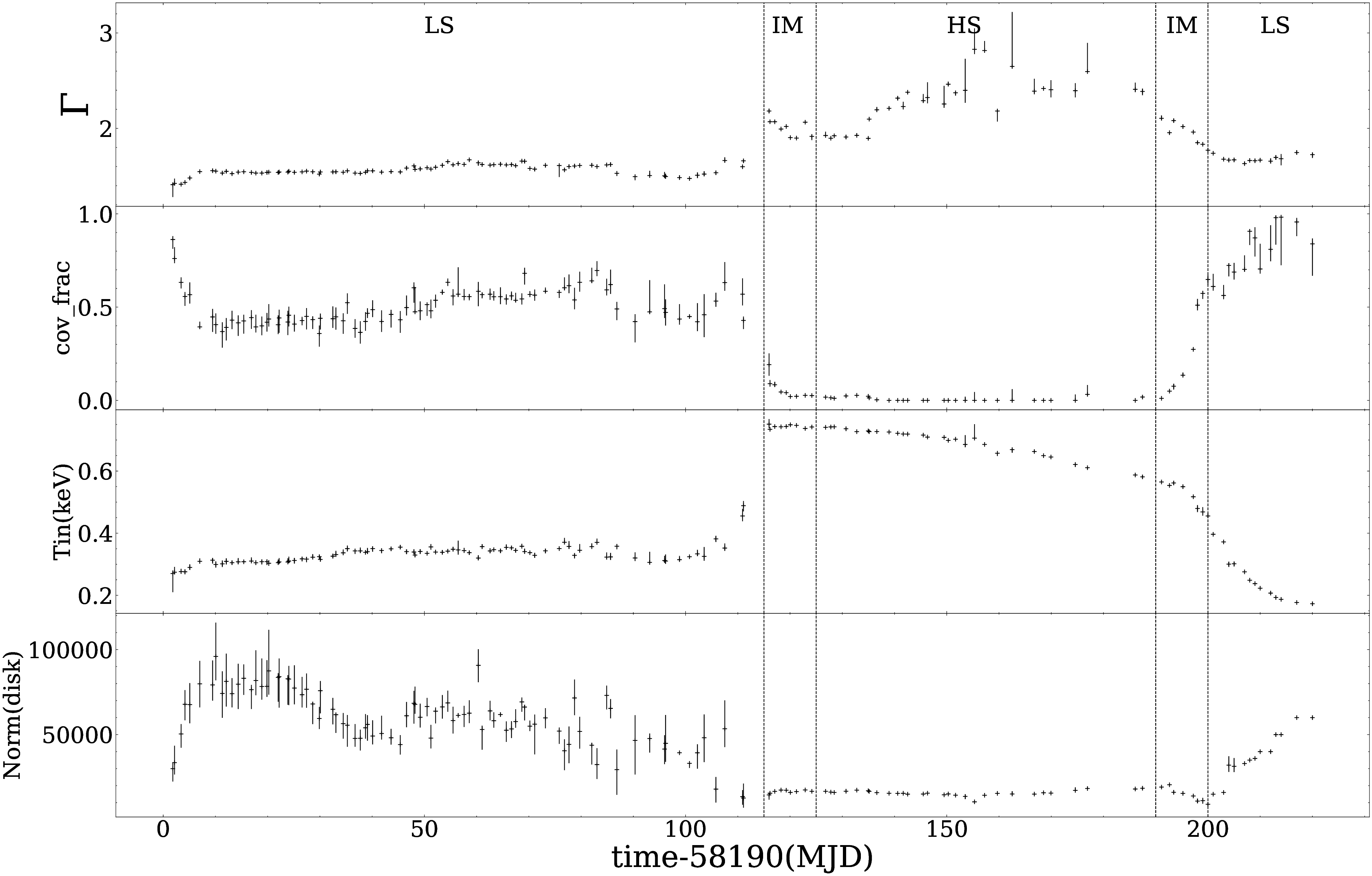}
	\caption{Spectral evolutions of the  outburst:  the low energy power-law photon index, Cov-frac  the coverage factor, \textit{T}in  the temperature of the inner disk and Norm  the normalization of the disk.}
	\label{spectrum}
\end{figure*}

\begin{figure}
	\centering
	\includegraphics[angle=0,scale=0.17]{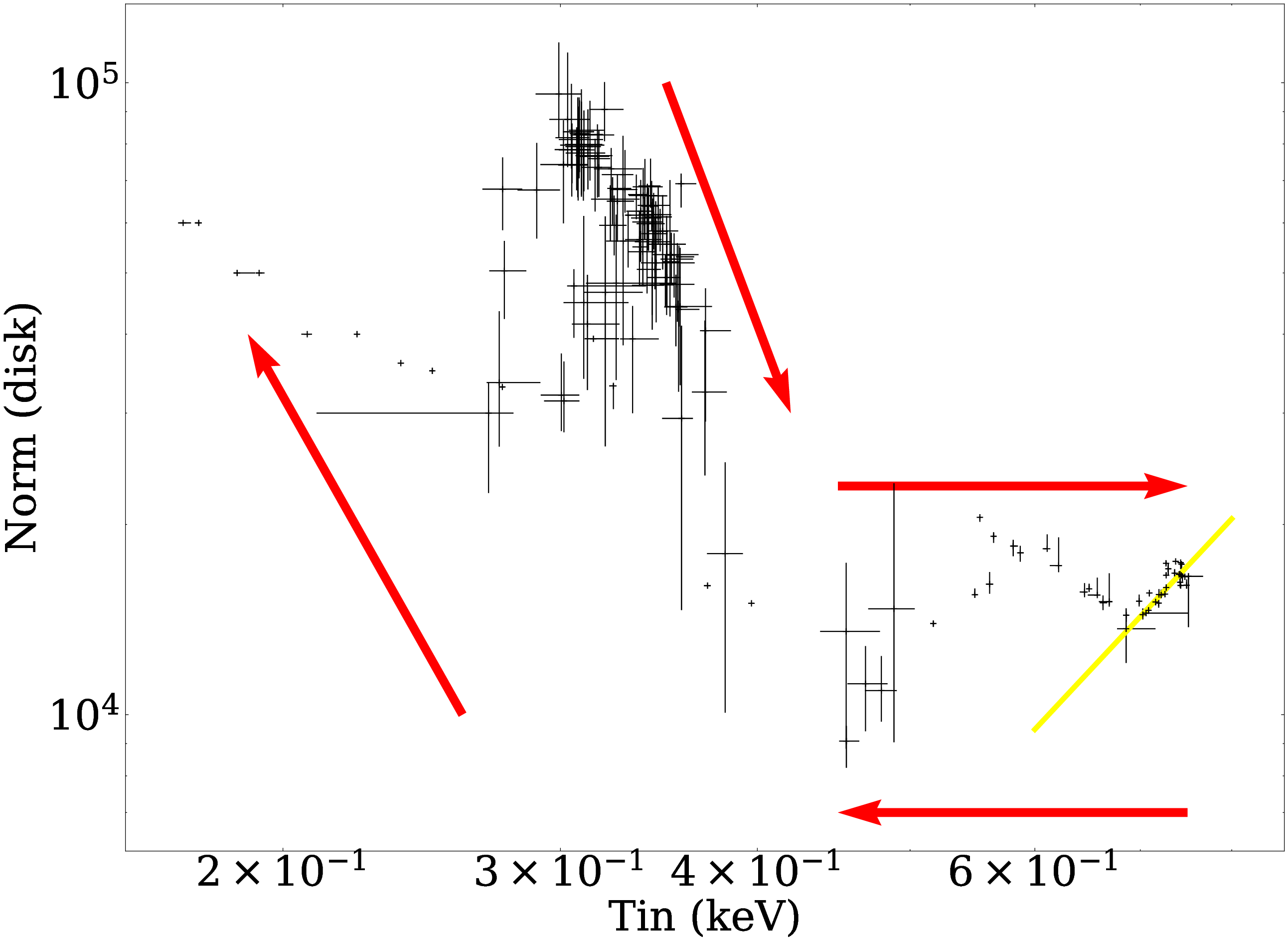}
	\caption{  The evolution of the normalization and the temperature of the inner disk. The red arrows indicate the evolution direction. The yellow line with a slope of 8/3.}
	\label{RT}
\end{figure}

\textit{Insight}-HXMT has a wide energy band, which helps distinguish various spectral components and study the evolution of spectral parameters. The spectral fitting is carried out using the software package {\tt{XSPEC V12.12.1}}. We  adopt several models to fit the spectrum of \textit{Insight}-HXMT throughout the whole outburst from MJD 58191 to MJD 58412. 

We use TBABS model components to consider the interstellar absorption effect \citep{2000Wilms}. Because of the calibration issue of LE, we ignore photons below 2 keV,  the column density $ N\rm _H$ can not be well constrained and hence fixed  at 1.5 x $10^{22}$ $\rm cm^{-2}$ \citep{2018Uttley}. 
We multiply a CONSTANT component to account for the calibration discrepancies between \textit{Insight}-HXMT LE, ME and HE (in this paper, we fix the constant of LE to 1). 
DISKBB model is used to fit the multi-temperature blackbody component of the accretion disk \citep{1984Mitsuda}.
We convolve the THCOMP model with the DISKBB model to fit the thermally comptonized continuum formed by the Comptonization  of   the thermal electrons \citep{2020Zdziarski}. 
At this time, there is an obvious reflection component in the residual ($\chi^2$/(d.o.f)=738/351=2.102).
So we add the relativistic reflection model RELXILLCP to fit the reflection component in the spectrum \citep{2016Dauser}. The RELXILLCP model is a kind of RELXILL model, which contains a comptonized continuum, so we link k$\rm T_e$ and $\Gamma$ in RELXILLCP to k$\rm T_e$ and $\Gamma$ in THCOMP. 
During fittings,  we fix the spin of the black hole at $a_*$ = 0.2 \citep{2021Guan}, the inclination of the accretion disk at \textit{i} = 63\textdegree  \citep{2020Torres}, the two emissivity indices  at 3, and the inner radius of the disk  at the innermost stable circular orbit (ISCO) \citep{You2021}. We note that the spectral fittings are not very sensitive to these parameters, e.g. if or not to set the inner radius to ISCO, the normalization of diskbb hardly  changes. The reflection fraction is fixed to -- 1, so we only consider the reflected emission. The ionization parameters, iron abundance, and normalization of the model are free.
After adding the reflection model, $\chi^2$/(d.o.f.)=405.68/348=1.16.
In the first interval of the  outburst, there is an obvious iron emission line seen in the spectral residual,  so we add a GAUSSIAN model to fit the iron  line and fix the line energy at 6.4 keV ($\chi^2$/(d.o.f.)= 357.05/346=1.03).
Therefore, our fitting model is: CONSTANT*TBABS*(THCOMP*DISKBB+GAUSSIAN+RELXILLCP). When  the source entered the later  hard state, no obvious iron line was visible, thus we removed the GAUSSIAN model.
 
After the source returned to the hard state, there is no obvious reflection component. There is no need to consider  the reflection model RELXILLCP. There may be two possibilities to account for the weak reflection.  First, at the end of the decay phase, the flux is low, and the statistics of the data are not enough to see the reflection component. Second, if the corona covers the disk, the reflection emission will be diluted from the re-procession in the corona.
Errors  are calculated at the 90 percent confidence level.  Figure \ref{spectrum1} shows the spectral fittings of three typical observations in  zones I, II and III,  respectively.

The evolution of spectral parameters is shown in Figure \ref{spectrum}.
In the LHS, the photon index is relatively low, $\Gamma$ $\sim$ 1.5, and increases slightly, while the photon index $\Gamma$ $\sim$  3 and descends to  $\sim$ 1.5 in HSS.
The coverage factor is close to 1 at the beginning of the outburst, and then gradually decreases  until MJD 58200. Then the coverage factor is stable between 0.4 and 0.5 from  MJD  58200  to  MJD  58240. After entering the soft state, the coverage factor is almost zero. During the decay phase of the  outburst, when the source returned to the hard state,  the coverage factor increases  to approach  1. The temperature of the inner disk increases slightly from $\sim$ 0.3 to $\sim$ 0.4 during the first outburst, and then jumps to around high $\sim$ 0.7, once the source enters the HSS in the consecutive outburst and gradually decreases afterward. 
The normalization of the disk emission has relatively large values in the first interval of the outburst when the source stayed in the low hard state, then drops  when the source was in the HSS. These results are  generally consistent with the previous reports specific to the spectral investigations of the hard state in the first interval of the outburst \citep{You2021} and of the soft state in the  outburst \citep{2021Guan}.
 As shown in Figure \ref{RT}, with the increase of mass accretion rate, the radius of the inner disk decreases with the increasing temperature, and then becomes stable around ISCO in HSS, and the distribution of these points in HSS is consistent with the yellow line with a slope of 8/3, corresponding to a constant disc accretion rate.  During the decay phase, once the source returns to LHS again,  the inner disk radius enlarges with the decreasing temperature. 
The evolution of spectral parameters is generally consistent with what is usually observed from other BHBs (see \cite{2006McClintock}).

\section{discussion and conclusion}
\label{dis}

\begin{figure*}
	\centering
	\includegraphics[angle=0,scale=0.4]{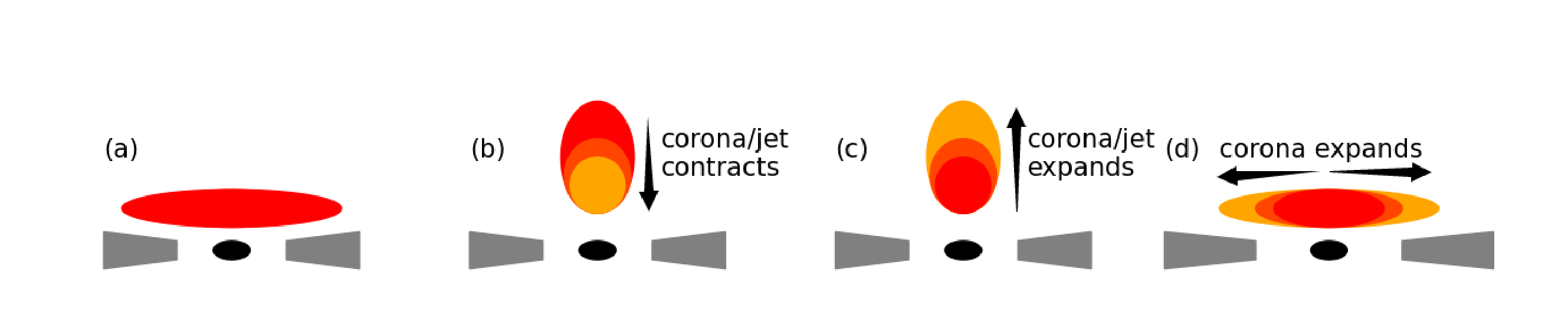}
	\caption{Schematic figure illustrating the  evolution of the corona and jet  of MAXI J1820+070. (a) At the beginning of the outburst, there is a radial corona covering the surface of the disk. (b) In the low hard state, the radial corona disappears and evolves into a vertical jet that contracts vertically. (c) In the hard-to-soft state transition, the jet expands. (d) In the low hard state, the corona expands radially and covers the surface of the disk.}
	\label{corona}
\end{figure*}

We have analyzed the  outburst of MAXI J1820+070 observed thoroughly by \textit{Insight}-HXMT. Following the procedure proposed by \cite{weng2021} on reshaping the HID diagram of MAXI J1348--630 from ‘q’ to a linear one by correcting the time lag between the hard and the soft X-rays, we find different time lags during the outburst of MAXI J1820+070. During the first interval of the outburst when the source was in the hard state and the high soft state of the second interval of the outburst, soft X-rays lag the hard X-rays by $3.5^{+0.02}_{-0.48}$ days and $4.93^{+0.72}_{-0.53}$  days, respectively. Such time lag reverses once the source steps into the decay phase of the  outburst and manifests as a hard X-ray lag of $10.82^{+1.32}_{-0.39}$ days. We apply the lag correction to the ‘q’ track experienced by the source in the HID and find that the tracks reform into a linear relationship for periods when the source stays in the hard states (e.g. the first interval of the outburst and the decay phase of the  outburst). However, the region of HSS in HID remains unchanged and thus appears insensitive to the time lag correction. These results confirm the report from \cite{weng2021} in their argument of the time lag correction in HID and provide more clues in understanding the ‘q’ shape in HID by addressing in additional the HSS and decay phase of the outburst.

\cite{weng2021} took a new view of the ‘q’ shape in HID by applying time lag correction between hard and soft X-rays. They found for the outburst of MAXI J1348--630, the optical and hard X-ray fluxes are linearly correlated and both lead the soft X-rays by about $10.07^{+0.69}_{-2.38}$ days. Accordingly, after correction of such a time lag between soft and hard X-rays, the traditional ‘q’ shape transforms into a linear one, which may support the following scenario of an outburst: The hard X-rays from the central corona heat up the outer part of the disk, leading to the contemporary optical and UV emission, then the fluctuation moves inward viscously until reaches to the inner disk where soft X-rays show up. Here for \textit{Insight}-HXMT observation of outburst of MAXI J1820+070, we derive similar results for the first interval of the outburst when the source was in the long-lasting hard state: the hard X-rays lead by $3.5^{+0.02}_{-0.48}$ days and ‘q’ shape can be reformed into a linear one after the lag correction.  We notice that in \cite{weng2021} the \textit{Insight}-HXMT observations covered the hard/intermediate spectral states but less the HSS and decay phase of MAXI J1348--630, while the latter was observed thoroughly by \textit{Insight}-HXMT for MAXI J1820+070. Accordingly, the new information added to this end is the soft time lag in HSS and hard time lag in decay phase, with which we speculate that they may be understood in an overall scenario of corona/jet evolution during outburst, as illustrated in what follows. 

NICER and \textit{Insight}-HXMT observations of the first interval of the outburst of MAXI J1820+070 revealed that  during the hard state most of the hard X-rays are from an evolving jet: along with the jet base sinking toward the BH \citep{Kara2019}, the jet is precessing to produce low-frequency QPO \citep{2021Ma} and the hot flow confined within the jet moves outward in an accelerated way \citep{You2021}. 
Such a jet is also supported in Figure \ref{frequency}: If the break frequency of the hard X-rays represents the fluctuations intrinsic to the corona/jet, then the closer the corona/jet to the central compact object the higher the break frequency. 
Also, it was reported from joint \textit{Insight}-HXMT, NuSTAR and INTEGRAL observations that during the hard state of the first interval of the  outburst, the inner disk is truncated \citep{Zdziarski2022}. Once the source steps into HSS in the  outburst, the inner disk is close to ISCO, and the BH spin is estimated via measuring the continuum thermal spectrum \citep{2021Guan, 2021Zhao}. Here we show similar properties of disk and corona/jet in our spectral analysis of the two intervals of the outburst: as shown in Figure \ref{spectrum}, the disk temperature remains low in the first interval of the  outburst and jumps to around 0.7 keV once the source enters the HSS in the  outburst; the disk normalization is higher in hard state than in HSS; the coverage fraction of the corona/jet remains around 0.5 in the hard state of the first interval of the outburst but decreases to around zero in HSS of the  outburst. These confirm that in the hard state of the  outburst, the jet is dominant while in the HSS the jet is replaced by a corona largely contracted to a relatively small central region so that its coverage to the disk emission is hardly visible. Moreover, the spectral analysis is supplemented by \textit{Insight}-HXMT observations of the start and end phases of the two consecutive outbursts. We see from Figure \ref{spectrum} that, at the early phase of the first interval of the outburst, the coverage fraction drops from $\sim$ 1 to less than 0.5 and increases from zero level to around unit during the ending decay phase of the  outburst. This may indicate a contracting corona and an expanding corona, respectively. 

In short, a possible overall scenario of corona evolution in the outburst of MAXI J1820+070 is illuminated in Figure \ref{corona} based on the results derived in this paper: At the beginning of the  outburst the corona covering the disk contracts and results in a decreasing coverage fraction; afterward, the corona is replaced by a precessing jet which has a coverage fraction around 0.5;  in HSS of the  outburst the jet turns to  a geometrically tiny corona with almost non-visible coverage fraction; in the decay phase of the outburst, a corona is born to gradually extend over the disk and has a coverage fraction increasing up to unity.
From a comparison of the outburst scenario as revealed in this paper for MAXI J1820+070 with that for GRS 1915+105  presented in \cite{2022Mendez}, we find that the episode proposed by \cite{2022Mendez} may appear for MAXI J1820+070 during its intermediate and high states, where a transition between jet and central corona is indicated, while at even lower flux levels (at the beginning and ending phases of the outburst), most likely the source accretion system should have an extended corona configuration.
Since the tiny corona in HSS may suffer little influence from the variability of the disk emission, lag correction turns out to fail in reforming the HSS data in HID into a linear relationship, otherwise one needs an evolving corona (may as well the jet base) accompanied with disk instability that is relevant to the outburst evolution to have a  linear correlation between luminosity and the hardness ratio after accounting for the time lags.
Regarding the reversal of the time lag between hard and soft X-rays in the decay phase of the  outburst, the lagging hard X-rays may correspond to the formation of the corona.

The flux returning to the low hard state in the decay phase of the BHXRB outburst is lower than the original level of the hard state. This is called hysteresis and is accompanied by the recovery of the corona \citep{2015Kylafis}.
Although so far still little is known about the nature of the hysteresis effect, it is thought to be highly related to the formation of the corona via e.g. the mechanisms of disk evaporation \citep{2000Meyer, 2001Meyer} and disk magnetic reconnection \citep{2012MNRASWilkins} etc. Whatever mechanisms it holds, the lagging hard X-rays should result  from a later-born corona in the decay phase rather than a corona cooling by the disk emissions in the proceeding hard state and HSS state. Again, as shown in Figure \ref{uncorrected HID}, a successful reformation of a linear relation in HID for data both in the proceeding hard state of the first interval of the outburst and in the decay phase of the  outburst, but with however the exclusion of HSS, reinforces that such a reformation of the ‘q’ shape in HID subjected to time-lag correction between hard and soft X-rays requires an evolving corona, and thus may provide a probe to infer the possible overall evolution of corona (probably as well jet or jet base) during an outburst.

\section*{Acknowledgements}

This work is supported by the National Key R\&D Program of China (2021YFA0718500), the National Natural Science Foundation of China under grants U1838201, U1838202,  U2038101, U1938103, 12273030, U1938107.
This work made use of data and software from the \textit{Insight}-HXMT mission, a project funded by the China National Space Administration (CNSA) and the Chinese Academy of Sciences(CAS). This work was partially supported by the International Partnership Program of the Chinese Academy of Sciences (Grant No.113111KYSB20190020).
This research has made use of  software provided by of data obtained from the High Energy Astrophysics Science Archive Research Center (HEASARC), provided by NASA’s Goddard Space Flight Center.

%%%%%%%%%%%%%%%%%%%%%%%%%%%%%%%%%%%%%%%%%%%%%%%%%%
\section*{Data Availability}
The data used in this paper can be available from the  website of \textit{Insight}-HXMT
(\url{http://hxmtweb.ihep.ac.cn/}).



\bibliographystyle{mnras}
\bibliography{ref} % if your bibtex file is called example.bib

%\begin{thebibliography}{99}
%\end{thebibliography}

%%%%%%%%%%%%%%%%%%%%%%%%%%%%%%%%%%%%%%%%%%%%%%%%%%

%%%%%%%%%%%%%%%%% APPENDICES %%%%%%%%%%%%%%%%%%%%%

%%%%%%%%%%%%%%%%%%%%%%%%%%%%%%%%%%%%%%%%%%%%%%%%%%

% Don't change these lines
\bsp	% typesetting comment
\label{lastpage}
\end{document}